%Paper: astro-ph/9509094
%From: TUROLLA@padova.infn.it
%Date: Tue, 19 Sep 1995 12:44:01 +0200 (WET-DST)
%&cp-aa
\def\MS{M_\odot}
\def\Teff{T_{\rm eff}}
\def\cs{c_{\rm s}}
%\refereelayout{}
\MAINTITLE{The observability of old neutron stars
accreting the interstellar medium}
\SUBTITLE{III. The solar proximity}
\AUTHOR{Silvia Zane $^1$, Luca Zampieri $^1$, Roberto Turolla $^2$
and Aldo Treves $^1$}
\INSTITUTE{$^1$ International School for Advanced Studies, Via Beirut 2--4,
34013 Trieste, Italy\newline
$^2$ Department of Physics, University of Padova, Via Marzolo 8, 35131
Padova, Italy}
\ABSTRACT{
Old isolated accreting neutron stars may show up among unidentified soft X--ray
sources detected by the ROSAT All Sky Survey.
We argue that the chances of
identification are greater for ONSs located in the closest overdense regions of
the solar neighbourhood. In particular, we consider the neutral hydrogen wall
in the second Galactic quadrant ($15^\circ < l < 150^\circ$) which shrinks the
estimated contour of the Local Bubble to $\approx 16-30$ pc. Due to their
vicinity, we expect $\sim 10$ ONSs to be detectable at a relatively high flux
limit ($\sim 0.1$ counts/s) in the 0.2--2.4 keV band.
This implies that about 5 \% of unidentified
sources above this threshold could be ONSs. No optical counterpart is present,
but EUV emission from these objects could be detected using EUVE Lex filter
at the highest sensitivity limits.}

\KEYWORDS{Stars: neutron--Ultraviolet: stars--X--rays: stars}

\THESAURUS{08.14.1--13.21.5--13.25.5}

\OFFPRINTS{zane\char'100 tsmi19.sissa.it}

\maketitle

\AUTHORRUNNINGHEAD{S. Zane, L. Zampieri, R. Turolla and A. Treves}

\MAINTITLERUNNINGHEAD{The observability of old neutron stars. III}

\titlea{Introduction}
The possibility that old, isolated neutron stars (ONSs) accreting the
interstellar material (ISM) may be detected as soft X--ray sources was
suggested
long ago by Ostriker, Rees \& Silk (1970). Such observations would be of the
greatest importance since they can provide a unique tool for investigating
the basic properties of NSs. Although the total number
of ONSs in the Galaxy could be as high as $10^9$, the detection of
these sources appears very difficult for two main reasons: their
intrinsic weakness, bolometric luminosity $\la
10^{31} \ {\rm erg\, s^{-1}}$ for typical ISM densities and star
velocities, and their emitted radiation, peaked around $\approx 50$ eV,
which falls in an energy band which is not easily accessible even to
spaceborne instrumentation.

The launch in the early 90's
of two satellites especially committed to the study of the extreme ultraviolet
and soft X--ray bands, EUVE and ROSAT, renewed the interest in accreting ONSs
and led a number of investigators to reconsider the problem of their
observability (Treves \& Colpi 1991; Blaes \& Madau 1993, hereafter TC and
BM respectively; Colpi, Campana \& Treves 1993; Madau \& Blaes 1994)
The main result is that, assuming
that a relic B--field $\sim 10^9$ G funnels the accretion flow onto polar
caps, thousands of ONSs could have been detected by ROSAT All
Sky Survey (R--ASS) and one possible ONS candidate in the Cyrrus Cloud has
been already identified by Stocke et al. (1995) in the {\it Einstein} MSS .

All these investigations relied on the assumption that the radiation spectrum
is a blackbody. In the case of polar
cap accretion $\Teff$ is about four
times larger than in the isotropic case,
if one assumes that the only effect of the magnetic field is to
reduce the emitting area. Detailed radiative transfer
calculations (Zampieri et al. 1995, hereafter ZTZT)
showed, however, that the emerging
spectrum may be sensibly harder; in
this case the majority of ONSs would emit a substantial fraction of
their flux above $\sim 0.3$ keV and should be detectable also by ROSAT PSPC.
Exploiting this result, Zane et al. (1995, paper I in the following)
have recently shown that the Galactic population of ONSs can contribute up to
25 \% of the unresolved X--ray excess detected by ROSAT in the 0.5--2 keV band.

Clearly the chance of detecting an ONS would be larger by far for sources
located in the solar neighbourhood. If we assume an ONSs spatial density
of $\sim 3\times 10^{-4} \ {\rm pc}^{-3}$ (BM, paper I), about
140 ONSs are present in a sphere of radius 50 pc centered on the Sun.
Unfortunately, the local interstellar medium (LISM) is underdense
and relatively hot so the advantage expected from the vicinity of the
source is deceptive because its bolometric luminosity is a factor $n_{\rm
LISM}/
n_{\rm ISM}\sim 0.07$ smaller (see section 3).
The LISM, however, is highly inhomogeneous and
contains at least one region, the Wall, where the gas density approaches
the average ISM density, $n\sim 1 \ {\rm cm}^{-3}$ (see Paresce 1984;
Diamond, Jewell \& Ponman 1995).

In this paper we consider
the detectability of accreting ONSs in the solar proximity,
focussing our attention on the closest overdense regions outside the Local
Bubble which surrounds the Sun. We suggest the intriguing possibility that
the nearest neutron stars may account for a non--negligible fraction
of the relatively bright unidentified R--ASS sources in the Wall direction.

\titlea{The local interstellar medium}
Although a relatively high number of ONSs, $\ga 100$, are expected
within $\sim 50$ pc from the Sun, their observability as
accretion powered sources is severely hindered by the shortage of
fuel. In fact, in
the presently accepted picture, the Sun is surrounded by a region,
the Local Bubble, where the plasma has both very low density ($n \sim
0.05-0.07 \, {\rm cm }^{-3}$) and high temperature ($T \ga 10^5 $ K).
In the scenario proposed by McKee \& Ostriker (1977), see also
Cox \& Anderson (1982), Cox (1983), the hot
gas would fill $\sim 70-80 \%$ of the interstellar space and a large
number ($\sim 2 \times 10^4$) of cool
($T \sim 80$ K), roughly spherical clouds are expected to be present.
Observational data support this model for the region beyond $\sim 50-100$
pc from the Sun (Knude 1979), but,
as discussed by Paresce (1984), soft X--ray, radio and color excess surveys
seem to indicate that no clouds are present at smaller distances and that
the denser material
is more probably organized into large, elongated, moving fronts within $\sim
50$ pc. The Sun itself is embedded in a medium, the Local Fluff,
which is warm ($T \approx 10^3-10^4$ K) and slightly overdense
($n \sim 0.1$ cm$^{-3}$) with respect to the Local Bubble on scales $\la
20$ pc (Diamond, Jewell \& Ponman 1995).

Because both the underdensity and the high temperature in the Local
Bubble work against the accretion process, the location of the
overdense, warm fronts in the local neighbourhood is
of great importance as far as the detection of ONSs is concerned.
The present picture indicates that the contour of the Local Bubble
in the Galactic plane is highly asymmetric, with four major discontinuities
in four different Galactic sectors (Paresce 1984).
In particular, a wall of neutral hydrogen is located very close to the
Sun in the second quadrant, $ 15^\circ < l < 120^\circ$.
According to Paresce (1984), the wall is roughly
parallel to the $l = 330^\circ-150^\circ$ axis and is located at
$d \leq 16$ pc, with an estimated depth of about 35 pc. The $N_{HI}$
contours presented by Frisch \& York (1983) are generally farther away,
with the denser material ($n \sim 1$ cm$^{-3}$) at $\sim 90$ pc from the Sun.
Welsh et al. (1994) have
derived a highly asymmetric contour of the Local Bubble
that in the second quadrant is roughly intermediate between those presented
by Paresce and Frisch \& York.
The minimum radius of the local cavity has been estimated to be
$\sim 25-30$ pc, but, as stressed by the same authors, their indirect method
could produce an underestimate of $N_{HI}$ at distances smaller than 50 pc.
A very recent analysis of ROSAT EUV data (Diamond, Jewell \& Ponman 1995)
has shown that $n$ reaches $\sim$ 1 cm$^{-3}$ at $\sim 25-30$ pc and this
result seems to be in agreement with the asymmetric contour found by Welsh
et al. more than with those of Paresce, Frisch \& York (see also
Pounds et al. 1993).
Despite this, the shrink of the local cavity to less than
$25-30$ pc from the Sun cannot be ruled out on the basis of present
observations.

Being very close, the overdense region in the second Galactic
quadrant (the Wall) provides one of the most favourable
environment for the detection of accreting ONSs. In the following sections
we will discuss this issue, using two limiting models for the LISM in the Wall:
in model I the overdense region is located between 16 and 50 pc, in model II
between 30 and 50 pc; in both cases the angular range is $15^\circ < l <
120^\circ$, $| b| < 30^\circ$.
If the total number of ONSs in the Galaxy and their local density are taken
to be $N = 10^9$ and $n_0 = 3 \times 10^{-4} \, {\rm pc}^{-3}$ respectively
(see paper I), we expect 22 (model I) and 18 (model II)
objects to be present in the Wall.
The observability of such objects depends on their
velocity distribution and on the physical parameters of the medium.
In order to account for the interstellar absorption, we
assume in model I a typical HI density
of $n = 0.1 \, {\rm cm}^{-3}$ in the Local Fluff and
$n = 1 \, {\rm cm}^{-3}$ between 16 and 50 pc (the Wall).
In model II we use $n = 0.1 \, {\rm cm}^{-3}$ within 20 pc,
$n = 0.07 \, {\rm cm}^{-3}$ for $20<d<30$ pc and $n = 1 \, {\rm cm}^{-3}$
for $30 < d < 50 $ pc. However, due to the high gas temperature in the Local
Bubble, interstellar absorption between 20 and 30 pc can be safely neglected.

Although X--ray surveys toward the Galactic
plane detect a very large number of sources, the extreme vicinity of these
objects could make them relatively bright and, as a consequence, detectable
at larger flux limits.

\titlea{Predicted number of sources}
The total bolometric luminosity emitted by an accreting neutron star depends on
both the star velocity and the density of the surrounding medium
and, for a NS with $M = 1.4 \MS$ and $R = 12.4 \ {\rm km}$, it is
given by
$$
L = 6.6\times 10^{31} \left({n \over {1\, {\rm H\, cm^{-3}}}}\right)
\left({v \over {{10}\, {\rm km\, s^{-1}}}}\right)^{-3} \ {\rm erg\, s^{-1}}\, .
\eqno(1)
$$
If the star moves subsonically relative to the ambient medium, the previous
formula is still valid provided that the local sound speed, $\cs$, is used
in place of the star velocity $v$.

The choice of the most favourable energy bands for detecting accreting ONSs
is crucially related to the spectral properties of the emitted radiation and
in particular on the mean photon energy. In the simplest assumption of
blackbody emission from the entire star surface, the spectrum is peaked at
$\sim 3 \Teff$, where
$$
\Teff = 3\times 10^5\left ( {L \over {10^{31}{\rm erg\, s^{-1}}}}
\right )^{1/4} \, {\rm K}\eqno(2)
$$
is the effective temperature. However, if the star retains a relic magnetic
field, the accretion flow can be channeled onto polar caps.
As a result of the smaller emitting area, the spectrum is harder (TC) and
the effective temperature is now
$$
\eqalign{
\Teff = & 1.2\times 10^6\left ( {L \over {10^{31}{\rm erg\, s^{-1}}}}
\right )^{1/4} \left ( {B \over 10^9 {\rm G }}\right )^{1/7}
\times \cr
& \left ( {v \over 10 \, {\rm km\, s^{-1}}} \right )^{3/14}
 \left ({1\, {\rm H cm^{-3}} \over n} \right )^{1/14}  \, {\rm K}\,;\cr }
\eqno(3)
$$
for typical parameter values $(\Teff)_{\rm mag}\sim 4(\Teff)_{\rm unmag}$.
Moreover, detailed radiative transfer calculations (ZTZT) which we have
already mentioned,
have recently shown that the actual spectrum sensibly deviates from a
blackbody at $\Teff$ and is harder by a factor 2--3 for $L\sim 10^{31} \
{\rm erg\, s}^{-1}$. This result is independent of the emitting area and is
due to the fact that
higher frequencies decouple at larger scattering depths in the NS atmosphere,
where temperature is higher: the emerging spectrum is a superposition of
blackbody spectra at different temperatures, which is broader than a
Planckian at $\Teff$.

For a typical value of the ISM density in the Local Bubble, $n=0.07 \ {\rm
cm}^{-3}$, and assuming $v=40 \ {\rm Km\, s}^{-1}$, the total
luminosity
is $\sim 7\times 10^{28} \ {\rm erg\, s}^{-1}$ and
these sources would be within the EUVE and ROSAT WFC bandpasses, regardless of
the details about the emitted spectrum. However, even if they are located at
a distance
of 20 pc, they are too faint to be above the sensitivity threshold of both
detectors. In the case of polar cap accretion, a
non--negligible fraction of the total luminosity is emitted in the 0.2--0.4
keV energy interval (S bandpass of ROSAT PSPC), producing a count rate
($\sim$ 0.1--0.6 counts/s) well above the sensitivity limit.
However, the expected number of these relatively bright nearby sources
is so small to be subject to large statistical fluctuations
and, in addition, it could be strongly affected by the
physical state of the gas in the Local Bubble which is poorly known:
if the temperature is as large as $10^6$ K, the accretion luminosity drops
below
$\sim 5\times 10^{27} \ {\rm erg\, s^{-1}}$ (corresponding to $v \simeq \cs
\simeq 100$ km/s) and these sources would become too faint.

On the other hand, if we restrict our attention to
the region of the Wall ($|b|< 30^\circ$, $15^\circ < l < 120^\circ$), where
$n\sim  1\, {\rm cm^{-3}}$, the luminosity is $\sim 10^{30} \
{\rm erg\, s}^{-1}$ and the mean photon energy ranges from
40 eV in the case of accretion on the entire NS surface
to 400 eV for polar cap accretion with a magnetic field of
$10^9$ G. In the following we will concentrate
only on the Wall direction:
as it will be discussed in detail later on, a number of
sources accreting in this locally overdense region turns out to be detectable
at least by some of these instruments.

Since the accretion luminosity depends on the star velocity, a statistical
approach is required to calculate the
expected number of detectable sources above a fixed sensitivity threshold.
In paper I, following BM, we derived
the present distribution function of ONSs in phase space, following the
evolution of
the orbits of $\sim 5\times 10^4$ stars taken as representative of the F
population considered by Narayan \& Ostriker (1991). These stars account for
$\sim 55$ \% of the total and correspond to fast rotating, low velocity
objects at birth. Here we are interested in investigating the
observability of these objects in the solar proximity, $d\la 50$ pc.
As already discussed in paper I,
the use of the analytical fit
$$
G(v) = {{\left ( v/v_0 \right )^m} \over { 1 +
{\left ( v/v_0 \right )^n}}} \eqno(4)
$$
to the computed cumulative velocity distribution is to be preferred whenever
an estimate of the number of ONSs
in a limited volume of phase space is needed; in the previous expression
$v_0 = 69 \, {\rm km \, s}^{-1}$ and $n\simeq m = 3.3$.
We assume that the star distribution is spatially homogeneous and use our
derived value for the local density, $n_0 =3\times 10^{-4} \, {\rm pc^{-3}}$.

The count rate measured at earth, corrected for the absorption
of the interstellar gas, is:
$$
CR = { 1 \over 4 \pi d^2}
\int_{\Delta E} {L_\nu \over {h \nu}} \exp \left({-\sigma_{\nu} N_{H}}
\right) A_\nu d \nu \eqno(5)
$$
where $A_\nu$ is the detector effective area,
$L_\nu$ is the monochromatic luminosity at the source,
$d$ is the distance, $N_{H}$ is
the column density
and $\sigma_{\nu}$ is the absorption cross section (Morrison \& McCammon 1983).
The effective areas were taken from Malina et al. (1994, Lex ASS), Edelstein,
Foster \& Bowyer (1995, Lex DE), Pounds et al. (1993, WFC)
and the ROSAT guide for observers (PSPC).

Three different spectral shapes have been considered: $a)$
blackbody emission from the entire surface, $b)$ blackbody emission
from the polar caps and $c)$ polar caps emission
using in this case the spectra calculated by ZTZT.
Once the shape of the emitted spectrum has been fixed, we
have calculated the
maximum value of $v$, $v_{\rm max}$, at which a star at the inner boundary
of the Wall, $d_1$, produces a count rate
above each chosen threshold.
For each value of $v$, the star distance was varied between $d_1$ and 50 pc
(the assumed outer boundary of the Wall for both models) in order to calculate
the maximum distance, $d_2 \left (v \right )$, at which such an object can be
detected in the Wall. The star will be observable within a volume $V \left (v
\right )$:
$$
V \left ( v \right ) = 2 \int_{l_1}^{l_2} d l \int_{\pi/3}^{\pi/2}
\sin \theta d \theta \int_{d_1}^{d_2} r^2 d r =
{\alpha \over 3 } \left ( d_2^3 - d_1^3
\right )\eqno(6)
$$
where $d_1 = 16$ pc (model I) and $d_1 = 30$ pc (model II),
$l_1 = 15^{\circ}$, $l_2= 120^{\circ}$ and
$\alpha = 1.83\ {\rm ster} \simeq 6000 \, {\rm deg^{2}}$
is the angular size of the Wall.
If $v_{\rm max}$ is larger then the sound
speed, the predicted number of sources is found integrating $dN/dv$ in
the range $0<v<v_{\rm max}$
$$
N = n_0 V \left ( \cs \right ) G \left (\cs \right ) +
n_0 \int_{\cs}^{v_{\rm max}} V \left (v \right ) {{dG}\over{dv}}\,
d v\, ;\eqno(7)
$$
the integral was evaluated numerically. In the accreting gas photoionized
by the star, $T \simeq 10^4$ K and $\cs\simeq$ 10 km/s.

Two distinct surveys, the All--Sky Survey (E--ASS) and the Deep Survey (E--DE)
were conducted with the EUVE telescopes.
Moreover observations with longer exposure times, the Right Angle
Program (RAP, McDonald et al. 1994),
allow the detection of sources with count rates down to $0.001$ counts/s, so we
repeated our calculations using also this limiting threshold.
We have considered the two filters
centered at higher frequencies, covering the wavelength range 58--364 A.
However, in the following
we will report only results for the Lex filter since our calculations in the
AlC bandpass indicate that no sources are expected to be observable in this
band.
As for the observability with ROSAT, we focussed on the total band T of PSPC
and WFC.
The bandpasses and limiting thresholds for these instruments are summarized
in table 1.

   \begtabfull
      \tabcap{1}{Bandpasses and thresholds for the EUVE and ROSAT instruments}
      \halign{#\hfil\quad&#\hfil\quad&#\hfil\quad&#\hfil\cr
            \noalign{\hrule}
            \noalign{\medskip}
            detector & filter & bandpass & threshold \cr
            & & (keV) & (ct/s) \cr
            \noalign{\medskip}
            \noalign{\hrule}
            \noalign{\medskip}
            EUVE ASS & Lex & 0.071--0.214 & 0.01 \cr
            EUVE DE & Lex & 0.069--0.183 & 0.015 \cr
            ROSAT WFC & S1 & 0.09--0.206 & 0.02 \cr
            ROSAT WFC & S2 & 0.062--0.11 & 0.025 \cr
            ROSAT PSPC & T & 0.2--2.4 & 0.015 \cr
            \noalign{\medskip}
            \noalign{\hrule}}
   \endtab

Above a sensitivity limit of $1.5 \times 10^{-2}$ counts/s
the R--ASS detected a very large
number of sources toward the Galactic plane.
If some ONSs are really present in the solar proximity,
we expect that their emission persists at larger flux limits where the number
of unidentified sources in the ROSAT survey is lower.
For this reason we have repeated our
calculations considering three different values for the threshold count rate:
$1.5 \times 10^{-2}$, 0.1 and 1 counts/s. Results are summarized in table 2
and 3.
The optical counterparts of ONSs are very faint (TC, BM), so the relatively
high count rate and the lack of optical identification of the X--ray
sources will be a distinguishing criterion.

   \begtabfull
      \tabcap{2}{Expected number of detectable sources for Model I}
      \halign{#\hfil\quad&#\hfil\quad&#\hfil\quad&$#$\hfil\quad&$#$\hfil\quad
              &$#$\hfil\cr
            \noalign{\hrule}
            \noalign{\medskip}
            detector & bandpass & threshold & N^{\rm a} & N^{\rm b} &
            N^{\rm c} \cr
            & & (ct/s) & & & \cr
            \noalign{\medskip}
            \noalign{\hrule}
            \noalign{\medskip}
            EUVE ASS & Lex & 0.01 & 2 & 7 & 1 \cr
            EUVE DE & Lex & 0.015 & 3 & 7 & 1 \cr
            EUVE RAP & Lex & 0.001 & 8 & 19 & 11 \cr
            ROSAT WFC & S1 & 0.02 & 1 & 3 & 0 \cr
            ROSAT WFC & S2 & 0.025 & 1 & 0 & 0 \cr
            ROSAT PSPC & T & 0.015 & 2 & 21 & 18 \cr
            ROSAT PSPC & T & 0.1 & 1 & 16 & 10 \cr
            ROSAT PSPC & T & 1. & 0 & 5 & 2 \cr
            \noalign{\medskip}
            \noalign{\hrule}}
      \halign{#\hfil \ \cr
            \noalign{\medskip}
            $^{\rm a}$ Blackbody emission from the entire star surface. \cr
            $^{\rm b}$ Blackbody emission from the polar caps, $B = 10^9$ G.
\cr
            $^{\rm c}$ ZTZT spectra, $B = 10^9$ G. \cr
            \noalign{\medskip}
            \noalign{\hrule}}
   \endtab

   \begtabfull
      \tabcap{3}{Same as in table 1 for Model II}
      \halign{#\hfil\quad&#\hfil\quad&#\hfil\quad&$#$\hfil\quad&$#$\hfil\quad
              &$#$\hfil\cr
            \noalign{\hrule}
            \noalign{\medskip}
            detector & bandpass & threshold & N^{\rm a} & N^{\rm b} &
            N^{\rm c}\cr
            & & (ct/s) & & & \cr
            \noalign{\medskip}
            \noalign{\hrule}
            \noalign{\medskip}
            EUVE ASS & Lex & 0.01 & 3 & 7 & 1 \cr
            EUVE DE & Lex & 0.015 & 4 & 8 & 1 \cr
            EUVE RAP & Lex & 0.001 & 9 & 17 & 12 \cr
            ROSAT WFC & S1 & 0.02 & 2 & 2 & 0 \cr
            ROSAT WFC & S2 & 0.025 & 2 & 0 & 0 \cr
            ROSAT PSPC & T & 0.015 & 2 & 17 & 16 \cr
            ROSAT PSPC & T  & 0.1 & 1 & 13 & 8 \cr
            ROSAT PSPC & T  &1. & 1 & 4 & 1 \cr
            \noalign{\medskip}
            \noalign{\hrule}}
   \endtab

The numbers in the tables suggest that, in the assumption of
blackbody emission, accreting ONSs could produce
a non--negligible count rate in the UV band.
Because the hardening of the spectra computed by ZTZT is comparable to that
induced by the presence of a magnetic field, we expect that, accounting
for more realistic spectral properties,
UV radiation could be detected even in the assumption of emission from
the entire NS surface. In this case
the number of detectable sources should be similar to the values of
N$^{\rm b}$ in tables 2--3.
The highest number of detectable objects corresponds to the intermediate
model in which only one source of
hardening acts, either the reduced emitting area or the differential
free--free absorption effect.
Since in this case the spectrum is peaked in the EUV--soft X--ray bands
and the LISM does not produce significant absorption at these
energies, the count rates turn out to be larger than those produced
by the ZTZT spectra emitted from the polar caps, which are
too hard to give a comparable contribution in the S band (0.2--0.4 keV) of
ROSAT PSPC.
However, we think that in a more plausible physical scenario
both a non zero magnetic field and the effects of bremsstrahlung opacity
should be accounted for.
Then, it follows that ONSs are, mainly, soft
X--ray emitters (see the last columns in tables 2--3), although the EUV
counterpart of very bright sources could be detected
by EUVE in the Lex band at the limiting sensitivity thresholds, 0.01 counts/s;
at the slightly higher sensitivity limit of $2 \times 10^{-2}$ counts/s,
the predicted number of sources is already zero.
In this respect
the RAP, improving the EUV sensitivity in pointed mode, seems to be the
most profitable way to search for the EUV counterparts of ONSs.
A detailed analysis of ROSAT PSPC ASS appears nevertheless the best approach
for detecting ONSs in the Wall. In particular, being such sources
very close, about 10 objects are expected to be observable with ROSAT PSPC
above a sensitivity limit of 0.1 counts/s.

\titlea {Comparison with present observations}
In order to compare our predicted number of sources with the actual
number of non optically identified sources (NOIDs)
observed so far in the direction of the Wall, we have
performed a systematic analysis of the ROSAT WFC ASS Bright Source
Catalogue (Pounds et al. 1993),
the First EUVE Source Catalog
(Bowyer et al. 1994), the EUVE Bright Source List (Malina
et al. 1994), the EUVE RAP source list (McDonald et al. 1994)
and the on--line catalogue of the ROSAT PSPC ASS
public pointings (White, Giommi \& Angelini 1994, WGA). We note that
R--ASS is not available for public consultations.

The WFC Bright Source Catalogue collects the observations
of the ROSAT WFC telescope, which carried out the first
almost complete survey
of the UV sky (96\%) in the 60--200 A
wavelength band. In addition,
the EUVE Bright Source List, the First EUVE Source Catalogue and the
EUVE RAP source list
contain the positive detections of sources in the E--ASS and E--DE.
About 97 \% of the sky has
been covered in E--ASS, while the deep survey spanned only a small strip along
the ecliptic plane.
As discussed in the previous section, relatively soft spectra
could produce a non--negligible EUV emission and it is therefore interesting
to analyze the present available data in this band. Searching in the
direction of the Wall, we found 5 sources without any
counterpart within a circle of $3$ \arcmin above $0.02$ counts/s in the EUVE
Lex filter and 5 unidentified sources in both the S1 and S2 WFC
filters; the EUVE RAP source list contain 4 new unidentified sources.
WFC sources  are not seen by EUVE. They are probably too
soft to be ONSs, because their S2 count rate always exceeds the S1 one, at
variance with what is expected for the majority of ONSs.
We stress, however, that it cannot be ruled out that some of
the faintest unidentified sources in E--ASS could be ONSs, if their emitted
spectrum is soft enough. In the case of blackbody
emission from the polar caps, we have calculated that 4 sources
can be detected in the Lex band above 0.02 counts/s which corresponds to 80
\% of EUVE NOIDs. However, the E--ASS is far from being complete at a
threshold of 0.02 counts/s ($\sim 3.5$ \% of the sky in the Lex band,
Bowyer et al. 1994) and the five detected sources are an absolute
lower limit for the total number of unidentified objects in the Wall.
As a consequence, the number of NOIDs in the Wall is consistent with
the expected number of ONSs in the EUV band.

The comparison with soft X--ray ROSAT observations has been performed on
the basis of the WGA catalogue, since R--ASS is not available for public
use. We found that pointings in the WGA catalogue cover
about 7\% (414 deg$^{-2}$) of the Wall, with
a total number of $\sim$ 7000 detected sources.
The number of objects
observed within an offset angle $\leq 20^{'}$ from the image center (where the
sensitivity of the detector is maximum) is given in table 4 as a function of
threshold.

   \begtabfull
      \tabcap{4}{WGA sources in the Wall direction
                 as a function of threshold}
      \halign{#\hfil\quad&#\hfil\quad&#\hfil\quad&#\hfil\quad&#\hfil\cr
            \noalign{\hrule}
            \noalign{\medskip}
            threshold & sources & source & expected & expected \cr
            & & density &
            sources$^{\rm a}$ & NOIDs$^{\rm a}$ \cr
            (ct/s) & & (deg$^{-2}$) & & \cr
            \noalign{\medskip}
            \noalign{\hrule}
            \noalign{\medskip}
            0.015 & 469 & 3.25 & 19500 & 1500 \cr
            0.1 & 132 & 0.91 & 5500 & 170 \cr
            1.0 & 48 & 0.33 & 2000 & 30 \cr
            \noalign{\medskip}
            \noalign{\hrule}}
      \halign{#\hfil \qquad \qquad \qquad \qquad \ \cr
            \noalign{\medskip}
            $^{\rm a}$ density $\times$ total area of the Wall. \cr
            \noalign{\medskip}
            \noalign{\hrule}}
   \endtab

We note that, quite independently of the assumptions
on the emitted spectrum, the total number of NOIDs is substantially
larger than our estimated number of observable ONSs in the Wall.
In particular, at a threshold of 0.1 counts/s, about
5\% of NOIDs could be ONSs, when the emitted
spectrum falls mainly within the ROSAT band. Clearly the number of NOIDs
decreases with increasing flux limit, so the search for ONSs could
conveniently be
restricted to sources above a relatively high ROSAT threshold, $\sim 0.1$
counts/s.
Although a slightly larger ONSs/NOIDs ratio is expected at higher flux
limits ($\sim 1$ count/s), the search for accreting NSs among such bright
sources could be fruitless because the estimated number is
so close to unity to be seriously biased by the uncertainties of the model.
On the other hand, the choice of a lower sensitivity limit, $\sim 0.01$
counts/s, does not provide a larger ONSs/NOIDs ratio, suggesting that
$0.1$ counts/s is indeed the most favourable threshold for identifying
nearby accreting ONSs.

\titlea{Conclusions}

The search for relatively bright R--ASS sources in
the Wall direction could be a promising strategy for selecting ONSs
candidates. At present this search is reserved to the groups which
have access to the R--ASS.
We have shown that old neutron stars may account for
a few percent of still unidentified sources above a threshold of
0.1 ct/s. In the case of polar cap accretion with the spectrum calculated by
ZTZT, which seems the more realistic assumption, the emitted
radiation is hard enough to give no detectable flux
in the visual band, $m_V\ga 29$. The absence of an optical counterpart
would be, therefore, a primary identification criterion. The R--ASS
candidates could be then searched for by E--DE in individual pointings. If
these objects have some radio emission, as suggested by Treves, Colpi \&
Lipunov (1993), another distinguishing feature would be the high proper motion,
$\sim 0.2 $ arcsec/yr assuming a velocity of 35 km/s and a distance of 30 pc.

\acknow{We thank Stuart Bowyer for helpful discussions
and the referee, Roger Malina, for bringing to our attention some recent
EUVE data.}

\begref{References}
\ref Blaes, O., Madau, P. 1993, ApJ, 403, 690 (BM)
\ref Bowyer, S., et al. 1994, ApJS, 93, 569
\ref Colpi, M., Campana, S., Treves, A. 1993, A\&A, 278, 161
\ref Cox, D.P., Anderson, P.R. 1982, ApJ, 253, 268
\ref Cox, D.P. 1983, in {\it Supernova Remnants and Their X--ray Emission\/},
IAU Symp. 101, eds. Danziger, J. \& Gorenstein, P., Reidel, Dordrecht
\ref Diamond, C.J., Jewell, S.J., Ponman, T.J. 1995, MNRAS, 274, 589
\ref Edelstein, J., Foster, R.S., \& Bowyer, S. 1995, ApJ, in press
\ref Frisch, P.C., York, D.G. 1983, ApJ, 271, L59
\ref Knude, J.K. 1979, A\&A Suppl., 38, 407
\ref Madau, P., Blaes, O. 1994, ApJ, 423, 748
\ref Malina, R.F., et al. 1994, AJ, 107, 751
\ref McKee, C., Ostriker, J.P. 1977, ApJ, 218, 148
\ref McDonald, K., et al. 1994, AJ, 108, 1843
\ref Morrison, R., McCammon, D. 1983, ApJ, 270, 119
\ref Narayan, R., Ostriker, J.P. 1990, ApJ, 270, 119
\ref Ostriker, J.P., Rees, M.J., Silk, J. 1970, Astrophys. Letters, 6,
179
\ref Paresce, F. 1984, AJ, 89, 1022
\ref Pounds, K.A., et al. 1993, MNRAS, 260, 77
\ref Stocke, J.T., et al. 1995, AJ, 109, 1199
\ref Treves, A., Colpi, M. 1991, A\&A, 241, 107 (TC)
\ref Treves, A., Colpi, M., Lipunov, V.M. 1993, A\&A, 269, 319
\ref Welsh, B.Y., Craig, N., Vedder, P.W., Vallerga, J.V. 1994, ApJ,
437, 638
\ref White, N.E., Giommi, P., Angelini, L. 1994, IAU Circ. 6100
\ref Zampieri, L., Turolla R., Zane S., Treves, A. 1995, ApJ, 439, 849
(ZTZT)
\ref Zane, S., Turolla, R., Zampieri, L., Colpi, M., Treves, A. 1995,
ApJ, in press (paper I)
\endref
\bye